\documentclass[11pt,a4paper]{article}
\pdfoutput=1
\usepackage{jheppub}
\usepackage{comment}
\usepackage{gensymb}
\usepackage{longtable}
\usepackage{amssymb}
\usepackage{lipsum}
\DeclareSymbolFont{matha}{OML}{txmi}{m}{it}
\DeclareMathSymbol{\varv}{\mathord}{matha}{118}
\usepackage{amsmath}
\usepackage{amssymb}
\usepackage{graphicx}
\usepackage{subcaption}
\usepackage{pstricks}
\usepackage{bm}
\usepackage{lscape}
\usepackage{pbox}
\usepackage{placeins}
\usepackage{booktabs}
\usepackage[T1]{fontenc}
\usepackage{footnote}
\usepackage{pdfpages}
\usepackage{hhline}
\usepackage{multirow}
\usepackage{multicol}
\usepackage{enumitem}
\usepackage[toc,page]{appendix}
\allowdisplaybreaks
\usepackage{comment}
\usepackage{float}                    
\usepackage{slashed}
\usepackage{mathrsfs}

\usepackage{color, colortbl}
\definecolor{LightCyan}{rgb}{0.88,1,1}

\usepackage{soul}
\setstcolor{red}
\usepackage{textcomp}
\usepackage{gensymb}
\usepackage{xcolor,pifont}
\usepackage{subcaption}

\newcommand*\colourcheck[1]{%
  \expandafter\newcommand\csname #1check\endcsname{\textcolor{#1}{\ding{52}}}%
}
\colourcheck{blue}
\colourcheck{green}
\colourcheck{red}
\colourcheck{olivegreen}
\definecolor{MyDarkBlue}{rgb}{0.1, 0.1, 0.8}

\usepackage{empheq}
\usepackage[numbers]{natbib}
\usepackage{notoccite}

\usepackage{rotating}
\usepackage{tabularx}
\usepackage[labelfont=bf]{caption} 

\FloatBarrier

\renewcommand{\thetable}{\arabic{table}}

\newcommand{\beq}{\begin {equation}}  
\newcommand{\eeq}{\end   {equation}} 
\newcommand{\bea}{\begin {eqnarray}} 
\newcommand{\eea}{\end   {eqnarray}}  
\newcommand{\baa}{\begin {array}   } 
\newcommand{\eaa}{\end   {array}   }     
\newcommand{\bit}{\begin {itemize} }
\newcommand{\eit}{\end   {itemize} }
\newcommand{\be }{\begin {equation}} 
\newcommand{\ee }{\end   {equation}}

\definecolor{MyDarkBlue}{rgb}{0.1, 0.1, 0.8} 
\definecolor{SBlue}{rgb}{0.2, 0.4, 0.7} 
\definecolor{MyLightBlue}{rgb}{0.22,0.51,0.9}
\definecolor{MyGreen}{rgb}{0.0, 0.5, 0.3}
\definecolor{BrickRed}{rgb}{0.8, 0.25, 0.33}
\RequirePackage{hyperref}
\hypersetup{colorlinks, citecolor=BrickRed,linkcolor=MyDarkBlue, urlcolor=MyLightBlue}

\begin{document}
\preprint{MS-TP-25-38}
\title{\textbf{Scrutinizing the KNT model with vacuum stability conditions}}
 \author[a]{Tim Huesmann,}
 \author[a,b]{Michael Klasen,}
 \author[a]{and Vishnu P.K.}

\affiliation[a]{Institut f{\"u}r Theoretische Physik, Universit{\"a}t M{\"u}nster, Wilhelm-Klemm-Stra\ss{}e 9, 48149 M{\"u}nster, Germany}
\affiliation[b]{School of Physics, The University of New South Wales, Sydney NSW 2052, Australia}
 
\emailAdd{thuesman@uni-muenster.de, michael.klasen@uni-muenster.de,  vishnu.pk@uni-muenster.de}
\abstract{The Krauss-Nasri-Trodden (KNT) model provides a unified framework for addressing the smallness of neutrino masses (by a three-loop radiative mechanism) and the dark matter abundance (via thermal freeze-out)  simultaneously. In this work, we investigate the implications of renormalization group effects on the model's parameter space. To this end, we perform a Markov Chain Monte Carlo analysis to identify the viable regions of parameter space that is consistent with all the relevant experimental and theoretical constraints at low energies.
We show that a significant portion of  the low-energy viable region  is incompatible with the vacuum stability conditions once the renormalization group effects are taken into account.  Most of the remaining parameter space of the model can be probed in future charged lepton flavor violating experiments.
}

\maketitle

\setcounter{footnote}{0}

\section{Introduction}\label{SEC-01}
The observed smallness of neutrino masses and the evidence for dark matter (DM) in the universe constitute two of the major unresolved puzzles in particle physics. Among the various proposals that address these issues, models  of radiative neutrino mass mechanism with a stable DM candidate are  particularly interesting. In these models~\cite{Tao:1996vb, Krauss:2002px, Ma:2006km, Aoki:2008av, Ma:2008cu, Restrepo:2013aga, Gustafsson:2012vj, Fiaschi:2018rky, Esch:2018ccs}, the symmetry that protects the stability of the DM candidate also forbids the tree-level contributions to neutrino masses, thereby ensuring neutrino mass generation through a radiative mechanism.
Due to the loop suppression, the corresponding new physics lies relatively at a lower scale,  making it testable at various experimental setups.

The Krauss-Nasri-Trodden (KNT) model~\cite{Krauss:2002px} is one of the first models proposed in this category. In this setup, the Standard Model (SM) is augmented with three fermion singlets (denoted as $N_{{1,2,3}}$) and two singly charged scalars (denoted as $S^{\pm}_{{1,2}}$). The DM stability and the absence of tree-level contributions to neutrino masses are ensured through imposition of a $\mathcal{Z}_2$ symmetry in this model.
Under this symmetry, all the BSM states, except  $S_1^{\pm}$, transform non-trivially.
The non-zero contributions to neutrino masses arise at three-loop level in this model. The observed dark matter abundance can be explained via a thermal freeze-out mechanism. The phenomenology associated with this model has been studied in various contexts, which include lepton flavor violating process~\cite{Cepedello:2020lul,Seto:2022ebh},  leptogenesis~\cite{Seto:2022tow}, a strong electroweak phase transition~\cite{Ahriche:2013zwa}, and direct and indirect searches of DM~\cite{Ahriche:2013zwa}.

In order to be consistent with the neutrino oscillation data and the dark matter relic density constraint, the couplings associated with the new Yukawa interactions of the KNT model are required to be sizable. This leads to testable predictions in various experiments, in particular the lepton flavor violating process $\mu\to e \gamma$ at the MEG II experiment~\cite{Seto:2022ebh}. Such sizable Yukawa couplings also could induce significant running for the scalar quartic couplings, which in turn may affect the  stability of the vacuum. These effects have previously been noted in several models of fermion singlets: the SM with right-handed neutrinos~\cite{Elias-Miro:2011sqh,Rodejohann:2012px,Khan:2012zw,Chakrabortty:2012np,DelleRose:2015bms}, the two-Higgs-doublet model with right-handed neutrinos~\cite{Mummidi:2018nph}, the scotogenic model~\cite{Lindner:2016kqk}, and leptophilic DM~\cite{Seto:2025pzw}. In this work, we investigate such effects within the context of the KNT model.

We find that a significant portion of the low energy viable parameter space of the model is incompatible with the vacuum stability conditions at a scale below the highest mass scale of the model. The generic solution to save these parameter regions is to introduce new physics (beyond the KNT model) below a physical mass scale of the model, which is inconsistent from a theoretical standpoint. This implies that the renormalization group (RG) induced effects drastically restrict the viable parameter space of the model.

The paper is organized as follows. In the next section, we present an overview of the KNT model. In Section~\ref{SEC-Constraints}, we summarize the relevant constraints applicable to the model. In Section~\ref{SEC-Results}, we discuss our numerical analysis and main results.  The conclusion of our work is presented in Section~\ref{SEC-Conclusion}.

\section{Model overview}\label{SEC-Model}
The KNT model~\cite{Krauss:2002px} extends the SM particle sector with three fermion singlets (denoted as $N_{{1,2,3}}$) and two singly charged scalars (denoted as $S_{1,2}^{\pm}$). In addition, the model also includes a $\mathcal{Z}_2$ symmetry, under which the SM particles transform trivially. The quantum numbers of the BSM states under $SU(3)_C\times SU(2)_L\times U(1)_Y\times \mathcal{Z}_2$ are given below:
\begin{align}
    N_{{i}}\sim (1,1,0;-); \, S^{\pm}_{1/2}\sim (1,1,\pm 1;+/-).
\end{align}
The lightest of $N_{{i}}$ can be a viable candidate for dark matter. 

The most general renormalizable Yukawa Lagrangian of the model (including the bare mass terms) is given below:
\begin{align}
    \mathcal{L}_{\mathrm{Yuk}}=&Y_u \overline{Q_L}\tilde{H} u_R + Y_d \overline{Q_L}H d_R + Y_e \overline{L_L}H \ell_R  \notag \\
    & + \tilde{Y}L^T_{L}Ci\tau_2L_{L} S_1^{+} +Y \overline{N^c}S_2^{+}l_{R}+\frac{M}{2} N^TCN + h.c.,
    \label{YukLag}
\end{align}
Here, $Q_L=(u,d)_L^T$ ($L_L=(\nu,\ell)_L^T$) denotes the left-handed quark (lepton) doublet and $H$ represents the SM Higgs doublet (with $\tilde{H}\equiv i\tau_2 H^*$). In Eq.~\eqref{YukLag},  the first row corresponds to the SM Yukawa Lagrangian, whereas the second row denotes the new Yukawa interactions with coupling matrices $Y$ and $\tilde{Y}$. Here, $Y$ is an arbitrary complex matrix, whereas $\tilde{Y}$ is an antisymmetric  matrix.

The scalar potential of the model involving $\{H,S_{1}^{\pm},S_{2}^{\pm}\}$ is given by
\begin{align}
\mathcal{V}=&-\mu^2_H (H^{\dagger}H)+\mu^2_{S_1}(S_1^+S_1^-)+\mu^2_{S_2}(S_2^+S_2^-)+\frac{\lambda_H}{2}(H^{\dagger}H)^2+\frac{\lambda_{S_1}}{2}(S_1^+S_1^-)^2+\frac{\lambda_{S_2}}{2}(S_2^+S_2^-)^2\nonumber  \\
&+\lambda_{HS_1}(H^{\dagger}H)(S_1^+S_1^-)+\lambda_{HS_2}(H^{\dagger}H)(S_2^+S_2^-) +\lambda_{S_1S_2}(S_1^+S_1^-)(S_2^+S_2^-)\nonumber\\ &+\frac{1}{2}\left( \lambda'_{S_1S_2} (S_1^+S_2^-)^2 + h.c.\right).
\label{PotLag}
\end{align}
The squared masses of the scalar states $\{h,S_{1}^{\pm},S_{2}^{\pm}\}$ can be written as
\begin{align}
    m^2_h&=\lambda_Hv^2, \nonumber \\
    m^2_{S_1}&=\mu^2_{S_1}+\frac{1}{2}\lambda_{HS_1}v^2, \nonumber \\
    m^2_{S_2}&=\mu^2_{S_2}+\frac{1}{2}\lambda_{HS_2}v^2.
\end{align}
Here, $h$ represents the SM Higgs boson and $v\simeq 246$ GeV denotes the electroweak vacuum expectation value.

\begin{figure}
    \centering
    \includegraphics[width=0.55\textwidth]{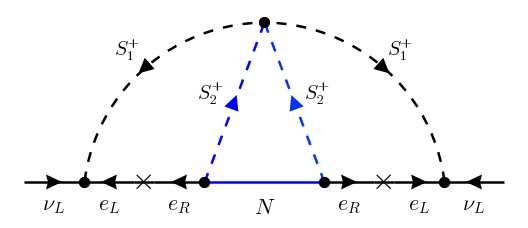}
    \caption{Neutrino mass generation in the KNT Model.}
    \label{fig:KNT_Mass}
\end{figure}

The neutrino masses and mixings are generated at three-loop order, see Fig.~\ref{fig:KNT_Mass} for the corresponding Feynman diagram. The resulting contribution to the neutrino mass matrix $M_\nu$ is given by~\cite{Ahriche:2013zwa}
\begin{align}
    M_\nu=c \tilde{Y}M_eY(M_{\textrm{eff}})^{-1}Y^TM_e\tilde{Y}^T.
\end{align}
Here, the factor $c=\frac{\lambda'_{S_1S_2} }{(4\pi^2)^3m_{S_2}}$ and the matrix $M_e=\textrm{diag}(m_e,m_\mu,m_\tau)$ corresponds to the charged lepton mass matrix. The effective matrix  $M_{\textrm{eff}}$ takes the following form
\begin{align}  (M_{\textrm{eff}})^{-1}=\textrm{diag}\Bigg(F\biggl(\frac{M^2_1}{m^2_{S_2}},\frac{m^2_{S_1}}{m^2_{S_2}}\biggr),F\biggl(\frac{M^2_2}{m^2_{S_2}},\frac{m^2_{S_1}}{m^2_{S_2}}\biggr),F\biggl(\frac{M^2_3}{m^2_{S_2}},\frac{m^2_{S_1}}{m^2_{S_2}}\biggr)\Bigg),
\end{align}
where the loop function $F$ is given by~\cite{Seto:2022ebh}
\begin{align}
    F(x,y)=\frac{\sqrt{x}}{8y^{\frac{3}{2}}}\int_{0}^\infty dr \frac{r}{r+x} \biggl(\int_{0}^1 dx \ln{\frac{x(1-x)r+(1-x)y+x}{x(1-x)r+x}}\biggr)^2.
    \label{eq:KNT_loop}
\end{align}

\section{Constraints}\label{SEC-Constraints}
Here, we discuss the relevant constraints applicable to the parameter space of the KNT model. We consider both theoretical and experimental constraints.   
 
\subsection{Theoretical constraints}
In this category, first we consider the constraints imposed by the vacuum stability conditions. For determining the implications of these conditions on the model parameters, we apply the co-positivity criteria discussed in Ref.~\cite{Kannike:2012pe}, which results in the following requirements on the scalar quartic couplings:
\begin{align}
&\lambda_H>0,\, \lambda_{S_1}>0,\, \lambda_{S_2}>0,\, \lambda_{HS_1}+\sqrt{\lambda_H\lambda_{S_1}} >0, \notag \\
&\lambda_{HS_2}+\sqrt{\lambda_H\lambda_{S_2}} >0,\, \lambda_{S_1S_2}-|\lambda'_{S_1S_2}|+\sqrt{\lambda_{S_1}\lambda_{S_2}}>0,\notag \\
&\sqrt{\lambda_H\lambda_{S_1}\lambda_{S_2}}+\lambda_{HS_1}\sqrt{\lambda_{S_2}}+\lambda_{HS_2}\sqrt{\lambda_{S_1}}+(\lambda_{S_1S_1}-|\lambda'_{S_1S_2}|)\sqrt{\lambda_H}  \notag \\ &+\sqrt{2(\lambda_{HS_1}+\sqrt{\lambda_H\lambda_{S_1}})(\lambda_{HS_2}+\sqrt{\lambda_H\lambda_{S_2}})(\lambda_{S_1S_2}-|\lambda'_{S_1S_2}|+\sqrt{\lambda_{S_1}\lambda_{S_2}})}>0 .
\label{eq:vacuumstab}
\end{align}

In addition, we also consider the constraints from the perturbativity conditions.  To satisfy these conditions, we demand the following conservative upper bounds on the scalar and the Yukawa couplings:
\begin{align}
|\tilde{Y}_{ij}|^2,|Y_{ij}|^2,|\lambda_{S_1}|,|\lambda_{S_2}|,|\lambda_{HS_1}|,|\lambda_{HS_2}|,|\lambda_{S_1S_2}|,|\lambda'_{S_1S_2}|\lesssim 4\pi.
\end{align}

\subsection{Experimental constraints}\label{subsec:expconst}
The relevant constraints in this category come from neutrino oscillation data, observed dark matter relic abundance, and the null results of several lepton flavor violation experiments. 

First, we consider the constraints imposed by neutrino oscillation data. To accommodate these constraints efficiently in our numerical analysis, we parameterize the Yukawa couplings $Y$ and $\tilde{Y}$ in terms of neutrino masses and mixings. The corresponding parametrization for the antisymmetric Yukawa matrix $\tilde{Y}$ is derived in Ref.~\cite{Cepedello:2020lul,Felkl:2021qdn}, which corresponds to
\begin{align}
     \frac{\tilde{Y}_{13}}{\tilde{Y}_{23}}&=\frac{c_{23}s_{12}}{c_{12}c_{13}}+e^{-i\delta_{CP}}\frac{s_{23}s_{13}}{c_{13}},\nonumber \\
    \frac{\tilde{Y}_{12}}{\tilde{Y}_{23}}&=\frac{s_{23}s_{12}}{c_{12}c_{13}}+e^{-i\delta_{CP}}\frac{c_{23}s_{13}}{c_{13}}
    \label{eq:anti-NH}
\end{align}
for the normal hierarchy (NH) and 
\begin{align}
     \frac{\tilde{Y}_{13}}{\tilde{Y}_{23}}&=e^{i\delta_{CP}}\frac{c_{23}c_{13}}{s_{13}},\nonumber \\
    \frac{\tilde{Y}_{12}}{\tilde{Y}_{23}}&=-e^{i\delta_{CP}}\frac{s_{23}c_{13}}{s_{13}}
    \label{eq:anti-IH}
\end{align}
for the inverted hierarchy (IH). Here   $c_{ij}\equiv \cos{\theta_{ij}}$, $s_{ij}\equiv \sin{\theta_{ij}}$,  $\theta_{ij}$ denotes the lepton mixing angles, and $\delta_{CP}$ stands for the Dirac CP phase. The above equations fix two of the three independent entries of $\tilde{Y}$. The remaining one is an arbitrary complex parameter, which without loss of generality is chosen to be $\tilde{Y}_{23}$. For determining the parameterization for $Y$, we closely follow the procedure given in Ref.~\cite{Cepedello:2020lul}. For this,  we rewrite the neutrino mass matrix in the following form:
\begin{align}
    M_\nu=-c\tilde{Y}M_\textrm{aux}\tilde{Y}
     \quad \mathrm{with} \quad M_\textrm{aux}=M_eY(M_{\textrm{eff}})^{-1}Y^TM_e.
    \label{eq:complexY}
\end{align}
Here, $M_\textrm{aux}$ is a complex symmetric matrix. Hence, it contains six independent complex entries, out of which three are arbitrary complex parameters. The remaining three can be expressed via Eq.~\eqref{eq:complexY} in terms of neutrino oscillation parameters, $\tilde{Y}_{23}$, and the three independent parameters of $M_\textrm{aux}$. The explicit forms of these expressions are  lengthy. Hence, we do not show them here. With  $M_\textrm{aux}=M_eY(M_{\textrm{eff}})^{-1}Y^TM_e$, the Yukawa matrix $Y$ can be expressed by using a modified version~\cite{Cepedello:2020lul} of Casas-Ibarra parametrization~\cite{Casas:2001sr} as
\begin{align}
    Y=M^{-1}_e U^T_\textrm{aux}(\hat{M}_\textrm{aux})^{\frac{1}{2}}R(M_\textrm{eff})^{\frac{1}{2}}.
\end{align}
Here, $\hat{M}_\textrm{aux}$ denotes the diagonal eigenvalue matrix of $M_\textrm{aux}$ and $U_\textrm{aux}$ represents the corresponding eigenvector matrix. We determine both these matrices numerically. The matrix $R$ is a  $3\times 3$ complex orthogonal matrix.

Next, we discuss the dark matter phenomenology associated with the KNT model. As mentioned earlier, the lightest of the $N_i$ can be a viable candidate for the dark matter in this model. Without loss of generality, we choose $N_1$ to be the dark matter state, which corresponds to the following assumption on the mass spectra of the BSM states:  $M_1<M_{2,3},m_{S_{1,2}}$. 
The observed dark matter abundance $ \Omega h^2\approx 0.12$~\cite{Planck:2018vyg} can be obtained via a thermal freeze-out mechanism.  The dominant processes that contribute to the DM annihilation cross section are $N_1N_1\to \bar{l}_il_j$,  which are induced by the Yukawa interactions $Y_{ij}\overline{N^c_{i}}S_2^{+}l_{R_j}$. Such contributions are $p$-wave suppressed. Consequently, the Yukawa couplings $Y_{ij}$ are required to be larger in order to be consistent with the dark matter relic density constraint. As we will show later, this requirement on the Yukawa couplings in effect induces significant running for both the Yukawa couplings and the scalar quartic coupling $\lambda_{S_2}$.  
In our analysis, we use {\tt{micrOMEGAs}}~\cite{Alguero:2023zol} to calculate the dark matter relic density. 
The constraints from DM direct or indirect detections experiments are typically relaxed for this scenario~\cite{Ahriche:2013zwa}. Hence, we do not consider them in our analysis.

The parameter space of the KNT model is also constrained stringently by lepton flavor violating (LFV) processes~\cite{Seto:2022ebh}.  The current limits on the relevant LFV processes are given below:
\begin{align}
    &\mathrm{BR}(\mu \rightarrow e\gamma)<3.1\times 10^{-13} \textrm{ \cite{MEGII:2023ltw}},  \\
       &\mathrm{BR}(\tau \rightarrow \mu\gamma)<4.2\times 10^{-8} \textrm{ \cite{Belle:2021ysv}}, \\
    &\mathrm{BR}(\tau \rightarrow e\gamma)<3.3\times 10^{-8} \textrm{ \cite{BaBar:2009hkt}},  \\ 
    &\mathrm{BR}(\mu \rightarrow 3e)<1.0\times 10^{-12} \textrm{ \cite{BELLGARDT19881}}, \\
    &\mathrm{CR}(\mu -e,Ti)<4.3\times 10^{-12}  \textrm{ \cite{DOHMEN1993631}}.
\end{align}
In our numerical analysis, we use  {\tt{SPheno}}~\cite{Porod:2011nf} to calculate the contributions to LFV processes.  

\section{Numerical analysis and results}\label{SEC-Results}
In this section, we discuss our numerical analysis and results. For our analysis, we use the following packages: {\tt{minimal-lagrangians}}~\cite{May:2020sod} is used to generate the input files for {\tt{SARAH}}~\cite{Staub:2015kfa}, which returns the analytical expressions for the RG equations and  also generates the input files for {\tt{SPheno}}~\cite{Porod:2011nf} and {\tt{micrOMEGAs}}~\cite{Alguero:2023zol}. As aforementioned, we use {\tt{micrOMEGAs}} to calculate the dark matter relic abundance, and we use {\tt{SPheno}} to calculate the contributions to various LFV processes.
\subsection{The viable parameter space at low energies}
Before investigating the implications of renormalization group effects on the parameter space of the KNT model, we first identify the regions that are consistent with  all the relevant constraints at low energies. For this, we perform a numerical scan over the free parameters in the KNT model. 

The model introduces 35 new real parameters, which include three fermion masses $M_{1,2,3}$, two scalar masses $m_{S_{1,2}}$,  six scalar couplings $\{\lambda_{S_1},\lambda_{S_2},\lambda_{HS_1},\lambda_{HS_2},\lambda_{S_1S_2},\lambda'_{S_1S_2}\}$, and twelve complex Yukawa couplings $\{Y_{ij},\tilde{Y}_{ij}\}$. 
The priors that we used  on the masses of the BSM states and the scalar quartic couplings in our numerical scan are given below:
\begin{align}
100\textrm{ GeV}<M_{1,2,3}&<10\textrm{ TeV}, \notag \\
100\textrm{ GeV}<m_{S_{1,2}}&<10\textrm{ TeV}, \notag \\
|\lambda_{S_1},\lambda_{S_2},\lambda_{HS_1},\lambda_{HS_2},\lambda_{S_1S_2},\lambda'_{S_1S_2}|&<4\pi. 
\end{align}
As for the Yukawa couplings, we parametrize them  in terms of neutrino masses and mixings, see Section~\ref{subsec:expconst} for the details. In our numerical scan, we allow for a $3\sigma$ range  given in Ref.~\cite{Esteban:2024eli} for the neutrino observables and consider both normal and inverted hierarchies for neutrino masses. The remaining free parameters Yukawa coupling $\tilde{Y}_{23}$, three angles $\alpha_i$ of the orthogonal matrix $R$, and three complex entries of the auxiliary matrix $M_{\rm aux}$ are allowed to be in the following ranges:
\begin{align}
 |\tilde{Y}_{23}|^2&<4\pi, \notag \\
    0<\alpha_i&<2\pi, \notag \\
    |(M_\textrm{aux})_{ij}|&<m_im_j\times(3\times1.05\times4\pi).
\end{align}
Here, the upper limit for the entries of $M_{\rm aux}$ is determined from their structure
given in Eq.~\eqref{eq:complexY} by considering
following upper bounds on $|Y_{ij}|<\sqrt{4\pi}$ and $F(x,y)<1.05$ \cite{Seto:2022ebh}.

To efficiently explore the multi-dimensional parameter space of the model, we perform a numerical scan using the affine-invariant Markov Chain Monte Carlo (MCMC) sampler implemented in the Python package \texttt{emcee}~\cite{Foreman_Mackey_2013}. As a preliminary step, we carry out a random scan to identify a sample point that satisfies all required theoretical and phenomenological constraints. 
From this sample point, we initialize multiple walkers to sample the surrounding regions. To reduce the dependency of the initial conditions, we define a burn-in phase of 2000 steps. We allow the sampler to run for $10^5$ iterations, which ensures sufficient coverage of the viable parameter space.  After constructing the resulting data set, we employ a normalizing flow algorithm to estimate the probability density of the viable points. Using this density estimate, we assign a weighting factor to each sample point, which is subsequently applied in the statistical analysis of the parameter space. 

\begin{figure}[ht]
    \centering
    \begin{subfigure}[b]{0.45\textwidth}
        \centering
        \includegraphics[width=\textwidth]{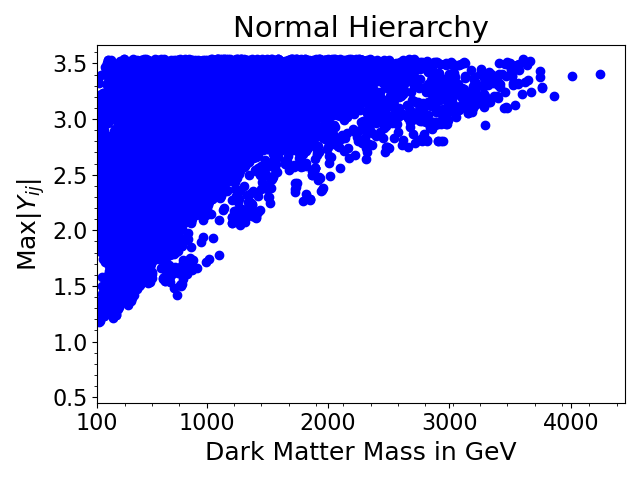}
    \end{subfigure}
    \begin{subfigure}[b]{0.45\textwidth}
        \centering
        \includegraphics[width=\textwidth]{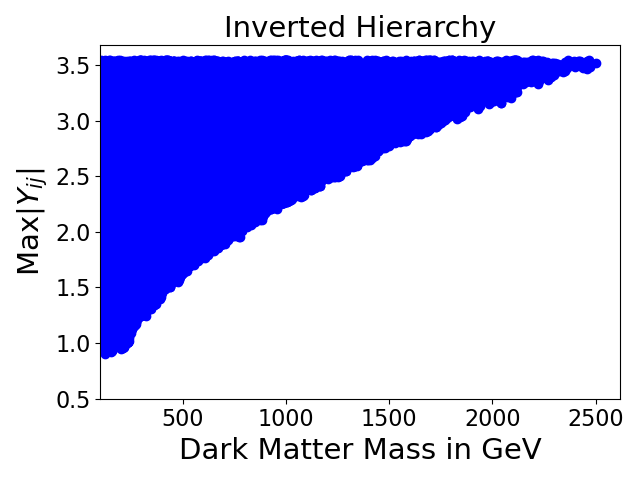}
    \end{subfigure}
    \caption{The parameter space in $\mathrm{Max}|Y_{ij}|$ versus dark matter mass plane consistent with all the relevant constraints at low energies for both normal (left panel) and inverted (right panel) hierarchies of neutrino masses.}
    \label{fig:parameter_space}
\end{figure}

In Fig.~\ref{fig:parameter_space}, we show the results of our numerical analysis in the plane of dark matter mass  versus  $\mathrm{Max}|Y_{ij}|$. It is important to notice that at least one of the entries of $Y$ is required to be of $\mathcal{O}(1)$. This is mainly because of the dark matter relic density constraint.  
As we will demonstrate in the next section, such sizable values of Yukawa couplings will induce significant running for both Yukawa couplings and scalar quartic couplings.

\subsection{Effects of the RG evolution}
We now investigate the implications of RG evolution of the couplings on the parameter space of the model. The relevant RG equations are computed using the package \texttt{SARAH}~\cite{Staub:2015kfa}, which are given in Appendix \ref{ch:RGE_KNT}. Using these RG equations, we evolve the couplings from the weak scale to a scale  where at least one of the theoretical constraints is violated for each of the low energy viable points. We then compare this scale of inconsistency, $\Lambda_\textrm{max}$, with the highest mass scale of the model, $M_\textrm{max}\equiv \mathrm{Max}\{M_{1,2,3},m_{S_{1,2}}\}$.

\begin{figure}[ht]
    \centering
    \begin{subfigure}[b]{0.45\textwidth}
        \centering
        \includegraphics[width=\textwidth]{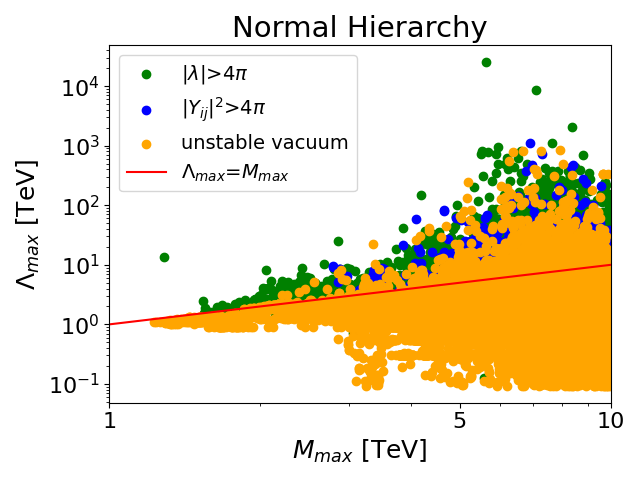}
    \end{subfigure}
    \begin{subfigure}[b]{0.45\textwidth}
        \centering
        \includegraphics[width=\textwidth]{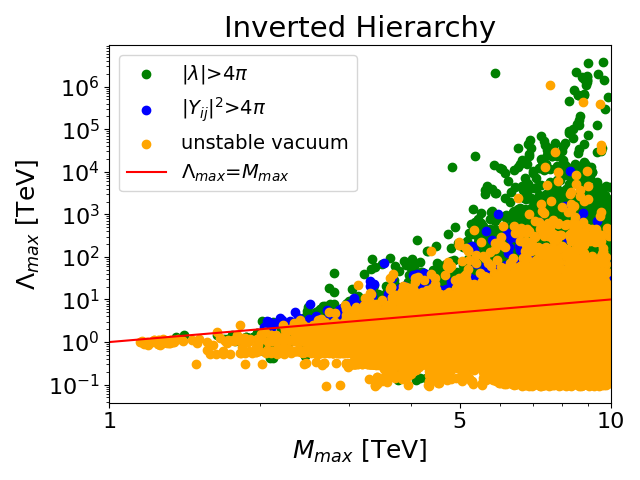}
    \end{subfigure}
    \caption{ 
    The scale of inconsistency ($\Lambda_\textrm{max}$) with respect to the heaviest mass scale ($M_\textrm{max}$) of the model for both normal (left panel) and inverted (right panel) hierarchies of neutrino masses.}
    \label{fig:Lambda_max}
\end{figure}

The resulting $\Lambda_\textrm{max}$ as a function of $M_\textrm{max}$ is shown in Fig.~\ref{fig:Lambda_max} for both NH (left panel) and IH (right panel). In this figure, the parameter points that violate the perturbativity conditions on the scalar quartic (Yukawa couplings) couplings at $\Lambda_\textrm{max}$ are  shown as green (blue) colored points. The orange colored points correspond to the case where the vacuum stability conditions fail at $\Lambda_\textrm{max}$. Notice that $\Lambda_\textrm{max}$ is not very high for these low energy viable solutions. Moreover, note that a majority of the parameter space lies below the $\Lambda_\textrm{max}=M_\textrm{max}$ line, see the first entry of Tab.~\ref{tab:consistency_scales} also.  This implies that only a small fraction of the low energy viable points remain consistent  with all the relevant constraints  up to the scale $M_\textrm{max}$. For the remaining parameter points, at least one of the theoretical conditions is violated below this scale and hence they should be discarded. To save these  parameter points, which are otherwise phenomenologically compatible, one could consider some additional new physics (beyond the KNT model) below $\Lambda_\textrm{max}$. However, such possibilities are theoretically inconsistent, since  $\Lambda_\textrm{max}$ is smaller than $M_\textrm{max}$ for these scenarios.

\begin{table}[htbp]
    \centering
    \begin{tabular}{||c|c|c||}
    \hline
       Consistency scale  &\multicolumn{2}{c||}{Fraction of parameter space}   \\
       \cline{2-3}
       & Normal Hierarchy  & Inverted Hierarchy \\
       \hline
        $M_\textrm{max}$ & 4.46\% & 9.23\% \\
        $2M_\textrm{max}$ & 1.15\% & 2.96\% \\
        $3M_\textrm{max}$ & 0.57\% & 1.66\% \\
        $4M_\textrm{max}$ & 0.30\% & 1.22\% \\
        $5M_\textrm{max}$ & 0.22\% & 0.99\% \\
        \hline
    \end{tabular}
    \caption{Fractions of the parameter space that are consistent up to a certain energy scale for both NH and IH.}
    \label{tab:consistency_scales}
\end{table}

\begin{figure}[ht]
    \centering
    \begin{subfigure}[b]{0.45\textwidth}
        \centering
        \includegraphics[width=\textwidth]{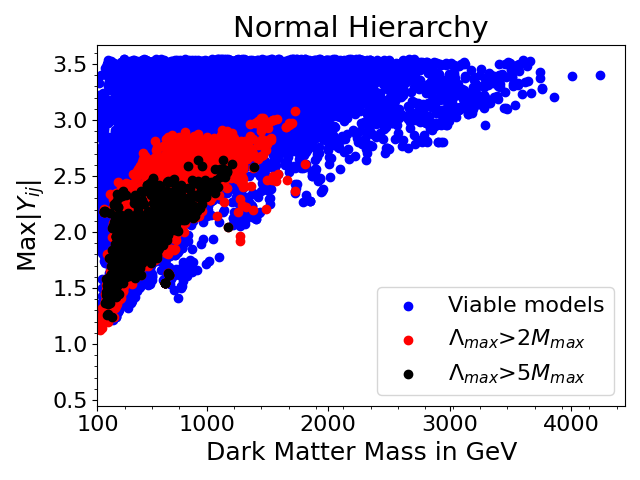}
    \end{subfigure}
    \begin{subfigure}[b]{0.45\textwidth}
        \centering
        \includegraphics[width=\textwidth]{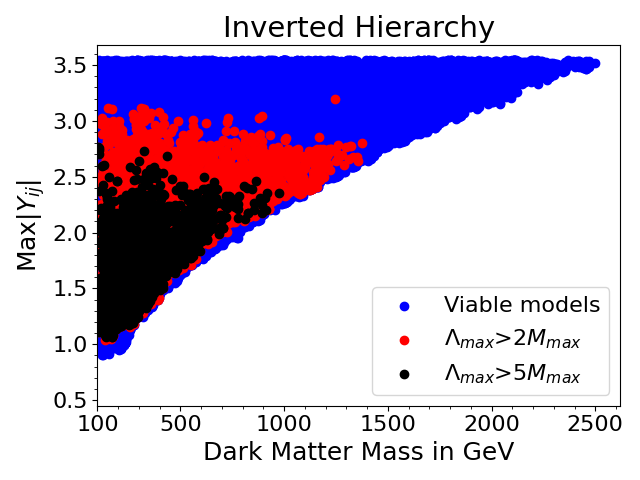}
    \end{subfigure}
    \caption{The parameter space in $\mathrm{Max}|Y_{ij}|$ versus dark matter mass plane consistent with all the relevant constraints at a scale above $2 M_{max}$ (red) and $5 M_{max}$ (black) for both normal (left panel) and inverted (right panel) hierarchies of neutrino masses.}
    \label{fig:parameter_space_modified}
\end{figure}

The RG induced effects on the model parameter space are even more stringent for the cases where the consistency is required to be above $M_\textrm{max}$.
To illustrate this, we consider various choices for the consistency scale in our analysis. The results  are given in Tab.~\ref{tab:consistency_scales} and Fig.~\ref{fig:parameter_space_modified} for both NH and IH. 
Notice that from Fig.~\ref{fig:parameter_space_modified}, the restriction induced by the RG effects is more severe for larger values of the Yukawa couplings. This suggests that these effects are mostly driven by the large entry of $Y$. 

\begin{table}[htbp]
    \centering
    \begin{tabular}{||c|c|c||}
    \hline
       Inconsistency scenario  & \multicolumn{2}{c||}{Fraction of parameter space}   \\
       \cline{2-3}
       & Normal Hierarchy  & Inverted Hierarchy \\ \hline
        Unstable vacuum & 85.46\% & 72.44\% \\
        $|\lambda|>4\pi$ & 12.31\% & 26.56\% \\
        $|Y_{ij}|^2>4\pi$ & 2.22\% & 1.00\% \\
        \hline
    \end{tabular}
    \caption{Distribution of the low energy viable points over the different scenarios for inconsistency for NH and IH.}
    \label{tab:inconsistency}
\end{table}

\begin{figure}[ht]
    \centering
    \begin{subfigure}[b]{0.45\textwidth}
        \centering
        \includegraphics[width=\textwidth]{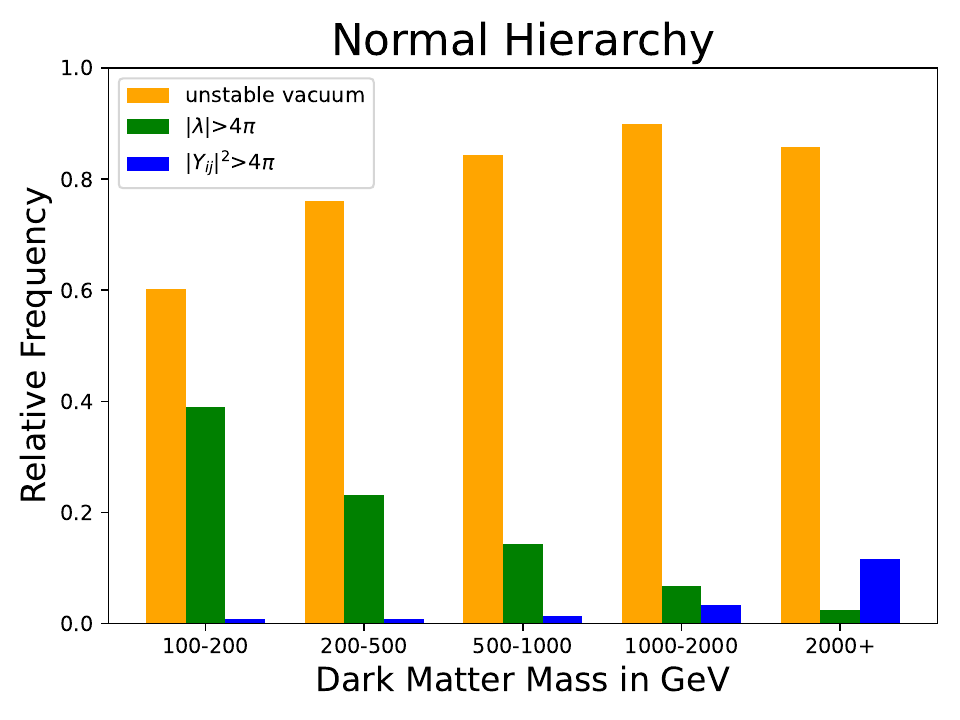}
    \end{subfigure}
    \begin{subfigure}[b]{0.45\textwidth}
        \centering
        \includegraphics[width=\textwidth]{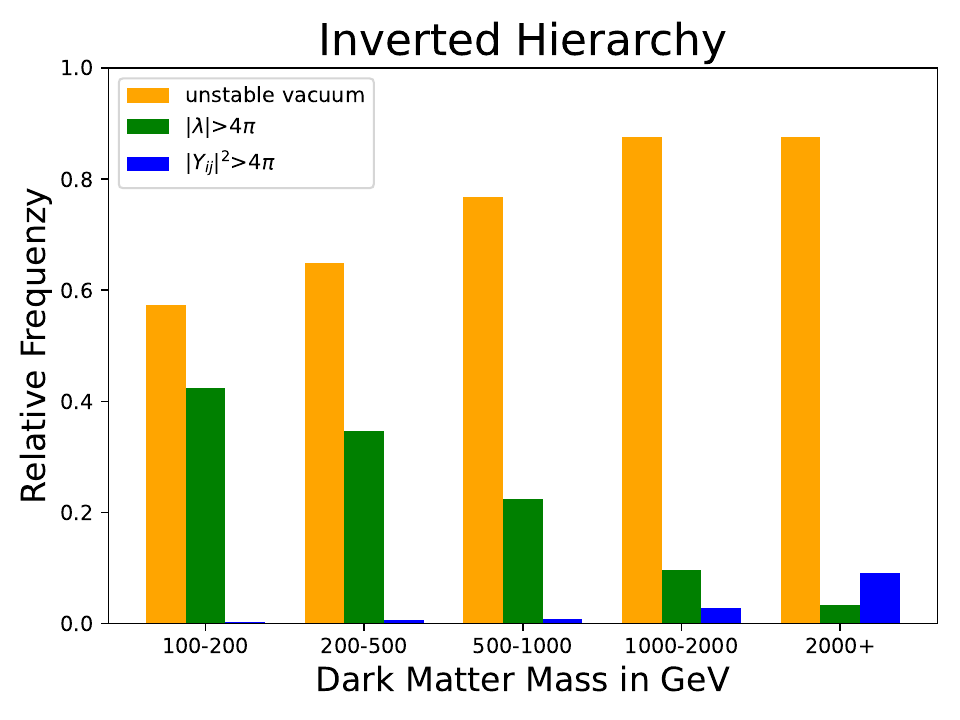}
    \end{subfigure}
    \caption{The relative frequencies of different theoretical requirements that fail at $\Lambda_{\mathrm{max}}$ as a function of dark matter mass for the case where $\Lambda_\textrm{max}\lesssim M_\textrm{max}$. }
    \label{fig:num_instability}
\end{figure}

\begin{figure}[ht]
    \centering
    \begin{subfigure}[b]{0.45\textwidth}
        \centering
        \includegraphics[width=\textwidth]{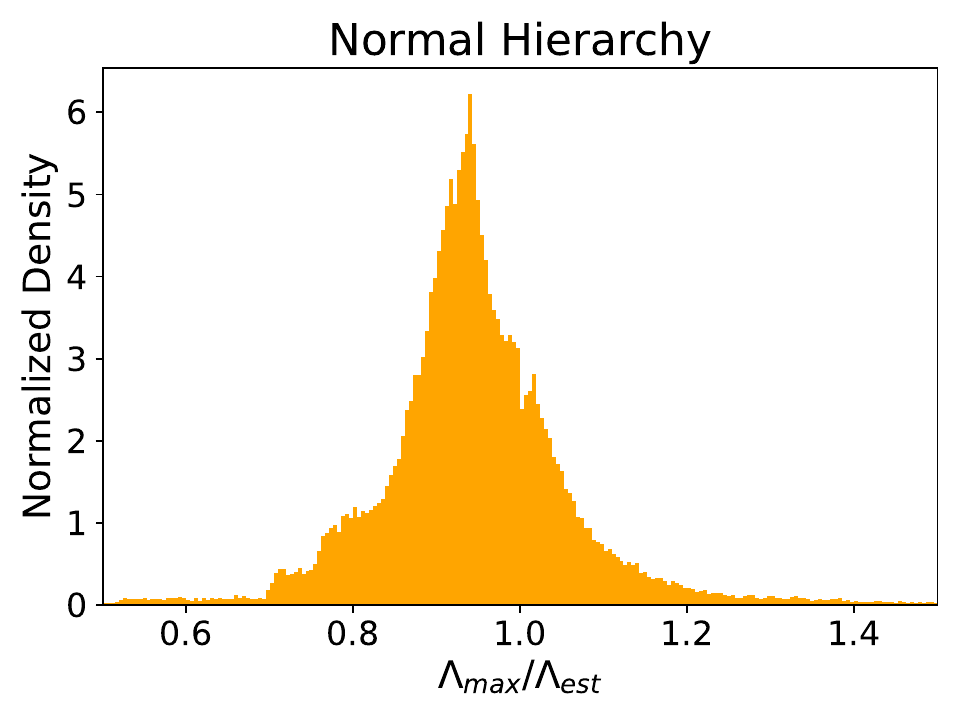}
    \end{subfigure}
    \begin{subfigure}[b]{0.45\textwidth}
        \centering
        \includegraphics[width=\textwidth]{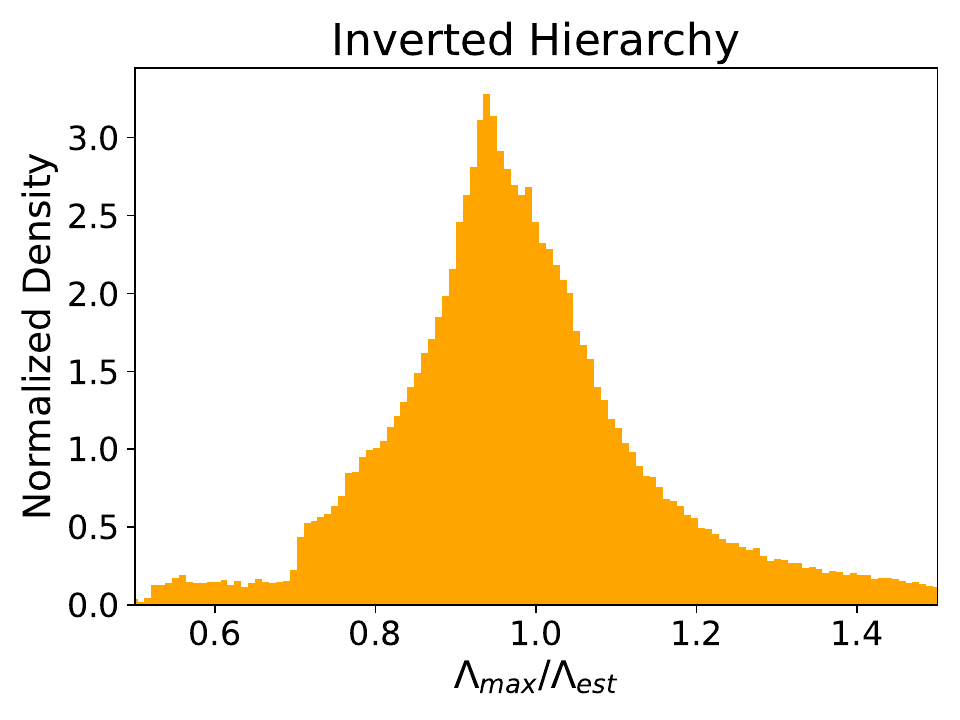}
    \end{subfigure}
    \caption{Distribution of the ratio $\Lambda_\textrm{max}/\Lambda_\textrm{est}$ shown as normalized density for both normal (left panel) and inverted (right panel) hierarchies of neutrino masses.}
    \label{fig:Lambda_est}
\end{figure}

In Tab.~\ref{tab:inconsistency}, we list the fraction of the low energy viable points that excluded by various theoretical constraints for the case where $\Lambda_\textrm{max}\lesssim M_\textrm{max}$. The results are shown in terms of relative frequencies as a function of dark matter mass in Fig.~\ref{fig:num_instability}. We find that the vacuum stability conditions impose the most stringent constraint on the model parameter space compare to that of perturbativity limits. Moreover, among the various vacuum stability conditions given in Eqs.~\eqref{eq:vacuumstab}, we find that the requirement $\lambda_{S_2}>0$ accounts for more than $99\%$ of the parameter points excluded by the vacuum stability conditions.
This clearly suggest that the RG contributions given in Eq.~\eqref{eq:betaS2} drive  $\lambda_{S_2}$ to negative values, consequently destabilizing the vacuum. Since we expect from Fig.~\ref{fig:parameter_space_modified} that these effects are  significant for larger values of  $Y$, the problematic term  can be identified with $\beta_{\lambda_{S_2}}\sim -4\mbox{Tr}(Y^\dagger YY^\dagger Y)$. To check whether this is the case, we estimate $\Lambda_\textrm{max}$ analytically (denoted as $\Lambda_\textrm{est}$) by only considering this term in $\beta_{\lambda_{S_2}}$, which yields  
\begin{align}
    \log\bigg(\frac{\Lambda_\textrm{est}}{\mu_0}\bigg)=\frac{4\pi^2\lambda_{S_2}}{\mbox{Tr}(YY^\dagger YY^\dagger)}\bigg|_{\mu=\mu_0},
\end{align}
where for simplicity we neglected the running of $Y$. Then, we computed $\Lambda_\textrm{est}$ and compared that with $\Lambda_\textrm{max}$ for the each of the low energy viable points that violates the vacuum stability conditions once the RG effects are taken into account. The results are shown in Fig.~\ref{fig:Lambda_est} for both NH and IH. As one can see that the normalized density has a peak at $\Lambda_\textrm{max}/\Lambda_\textrm{est}\sim 1$ for both scenarios, which implies that $\Lambda_\textrm{est}$ provides an accurate approximation of $\Lambda_\textrm{max}$.

\subsection{Future lepton flavor violating experiments}
We investigate the impact of future lepton flavor violating experiments on the remaining (small) fraction of viable points. The future sensitivity on various charged lepton flavor violating process are given below:
\begin{align}
    \mathrm{BR}(\mu\rightarrow e\gamma)&<6.0\times 10^{-14} \textrm{ \cite{MEGII:2023ltw}}, \\
    \mathrm{BR}(\tau\rightarrow e\gamma)&<3.0\times 10^{-9} \textrm{ \cite{Belle-II:2018jsg}},\\
    \mathrm{BR}(\tau\rightarrow \mu\gamma)&<10^{-9} \textrm{ \cite{Belle-II:2018jsg}}.
\end{align}
We find that about $90\%$ ($95\%$) of viable parameter space can be probed in future charged lepton flavor violating experiments for the scenario of NH (IH).

\section{Conclusion}\label{SEC-Conclusion}
In this work, we have investigated the impact of the RG effects on the parameter space of the KNT model. By performing a detailed numerical analysis, we have demonstrated that most of the low energy viable parameter points are incompatible with at least one of the theoretical constraints at a scale below the highest mass scale of the model. This in turn restrict the parameter space of the model significantly, since the generic solution, which is introducing new physics (beyond the KNT model) below a physical mass scale of the model  is inconsistent from a theoretical standpoint. 
Furthermore, we have also demonstrated that a vast region of the remaining parameter space, which amounts to above $90\%$, can be probed in future charged lepton flavor violating experiments.

Among the various theoretical constraints, we found that the vacuum stability conditions, in particular the requirement $\lambda_{S_2}>0$, impose the most stringent constraint on the parameter space of the model. The RG effects  driven by large values of the Yukawa couplings $Y$  lead $\lambda_{S_2}$ to negative values, consequently destabilizing the vacuum. To suppress these effects, one could consider tiny values for $Y$. However, such choices are not favored by the constraints imposed by the dark matter relic density and the neutrino oscillation data.

\section*{Acknowledgments}
M.K.\ thanks the School of Physics at the University of New South Wales in Sydney, Australia for its hospitality and financial support through the Gordon Godfrey visitors program. 


\appendix
\section{RG equations for the KNT Model}\label{ch:RGE_KNT}
In this section, we list the relevant 1-loop RG equations for the KNT Model. These equations are computed by using the package  {\tt{SARAH}}~\cite{Staub:2015kfa}. We write down these equations in terms of the beta functions, which are defined as 
\begin{align}
    \beta_g=16\pi^2\mu\frac{dg}{d\mu}
\end{align}
For convenience, we use the following short-hand notations
\begin{align}
    &T=\textrm{Tr}(Y^\dagger_eY_e+3Y^\dagger_uY_u+3Y^\dagger_dY_d) \\
    &T_{\tilde{Y}}=\textrm{Tr}(\tilde{Y}^\dagger \tilde{Y}) \\
    &T_Y=\textrm{Tr}(YY^\dagger) \\
    &T_4=\textrm{Tr}(Y^\dagger_eY_eY^\dagger_eY_e+3Y^\dagger_uY_uY^\dagger_uY_u + 3Y^\dagger_dY_dY^\dagger_dY_d) \\
    &T_{4\tilde{Y}}=\textrm{Tr}(\tilde{Y}^\dagger \tilde{Y} \tilde{Y}^\dagger \tilde{Y}) \\
    &T_{4Y}= \textrm{Tr}(YY^\dagger YY^\dagger)    \\
    &T_{\tilde{Y}e}=\textrm{Tr}(\tilde{Y}^\dagger \tilde{Y} Y^\dagger_eY_e) \\
    &T_{Ye}=\textrm{Tr}(Y^\dagger Y Y^\dagger_eY_e)
\end{align}
\subsection{Gauge couplings}
The gauge couplings associated with $U(1)_Y$, $SU(2)_L$, and $SU(3)_C$ symmetries are denoted by $g_1$, $g_2$, and $g_3$, respectively. The corresponding RG equations are given below:
\begin{align} 
\beta_{g_1} & =  
\frac{1}{3}\bigg(\frac{41}{2}+\theta(S^+_1)+\theta(S^+_2)\bigg) g_{1}^{3} \\  
\beta_{g_2} & =  
-\frac{19}{6} g_{2}^{3} \\ 
\beta_{g_3} & =  
-7 g_{3}^{3} 
\end{align}
\subsection{Yukawa couplings}
The Yukawa couplings of the model is defined in Eq.~\eqref{YukLag}. The beta functions associated with these couplings are given below:
\begin{align}
    \beta_{Y_u}&=Y_u\bigg(\frac{3}{2}(Y^\dagger_uY_u - Y^\dagger_d Y_d)+T-\frac{17}{12}g^2_1-\frac{9}{4}g^2_2-8g^2_3\bigg) \\
    \beta_{Y_d}&=Y_d\bigg(\frac{3}{2}(Y^\dagger_d Y_d -Y^\dagger_uY_u)+T-\frac{5}{12}g^2_1-\frac{9}{4}g^2_2-8g^2_3\bigg) \\
    \beta_{Y_e}&=Y_e\bigg(\frac{3}{2}Y^\dagger_eY_e -2\tilde{Y}^\dagger\tilde{Y}\bigg)+\frac{1}{2}YY^\dagger Y_e +Y_e\bigg(T-\frac{15}{4}g^2_1-\frac{9}{4}g^2_2\bigg) \\
    \beta_{\tilde{Y}}&= \tilde{Y}\bigg(4\tilde{Y}^\dagger \tilde{Y}+\frac{1}{2}Y^\dagger_eY_e \bigg) +\frac{1}{2}Y^T_eY^*_e \tilde{Y} +\tilde{Y}\bigg(4T_{\tilde{Y}}-\frac{3}{2}g^2_1-\frac{9}{2}g^2_2\bigg) \\
    \beta_Y&=YY^\dagger Y+Y_eY^\dagger_e Y+ Y\bigg(T_{Y} -\frac{9}{5}g^2_2\bigg)
\end{align}
\subsection{Quartic couplings of scalars}
The scalar quartic couplings are defined in Eq.~\eqref{PotLag}. The RG equations of these couplings are given below
\begin{align}
&\beta_{\lambda_H}=12\lambda^2_H+2\lambda^2_{HS_1}+2\lambda^2_{HS_2}+\frac{3}{4}\bigg(g^4_1+2g^2_1g^2_2+3g^4_2\bigg) +\lambda_H\bigg(4T-3(g^2_1+3g^2_2)\bigg) \nonumber \\
 & \qquad \quad -4T_4 \\
&\beta_{\lambda_{S_1}}=10\lambda^2_{S_1}+4\lambda^2_{HS_1}+2\lambda^2_{S_1S_2}+2|\lambda'_{S_1S_2}|^2+12g^4_1+\lambda_{S_1}\bigg(16T_{\tilde{Y}}-12g^2_1\bigg)-64T_{4\tilde{Y}} \\
&\beta_{\lambda_{S_2}}=10\lambda^2_{S_2}+4\lambda^2_{HS_2}+2\lambda^2_{S_1S_2}+2|\lambda'_{S_1S_2}|^2+12g^4_1+\lambda_{S_2}\bigg(4T_{Y}-12g^2_1\bigg)-4T_{4Y} \label{eq:betaS2}\\
&\beta_{\lambda_{HS_1}}=4\lambda^2_{HS_1}+6\lambda_H\lambda_{HS_1}+4\lambda_{S_1}\lambda_{HS_1}+2\lambda_{S_1S_2}\lambda_{HS_2}+3g^4_1\nonumber \\
&\qquad \quad +\lambda_{HS_1}\bigg(2T+8T_{\tilde{Y}}-\frac{3}{2}(5g^2_1+3g^2_2)\bigg)-16T_{\tilde{Y}e} \\
&\beta_{\lambda_{HS_2}}=4\lambda^2_{HS_2}+6\lambda_H\lambda_{HS_2}+4\lambda_{S_2}\lambda_{HS_2}+2\lambda_{S_1S_2}\lambda_{HS_1}+3g^4_1\nonumber \\
&\qquad \quad +\lambda_{HS_2}\bigg(2T+2T_{Y}-\frac{3}{2}(5g^2_1+3g^2_2)\bigg)-16T_{Ye} \\
&\beta_{\lambda_{S_1S_2}}=4\lambda^2_{S_1S_2}+4\lambda_{S_1}\lambda_{S_1S_2}+4\lambda_{S_2}\lambda_{S_1S_2}+4\lambda_{HS_1}\lambda_{HS_2}+8|\lambda'_{S_1S_2}|^2+12g^4_1 \nonumber\\
&\qquad \quad +\lambda_{S_1S_2}\bigg(8T_{\tilde{Y}}+2T_Y-12g^2_1\bigg) \\
&\beta_{\lambda'_{S_1S_2}}=\lambda'_{S_1S_2}\bigg(2\lambda_{S_1}+2\lambda_{S_2}+8\lambda_{S_1S_2}+8T_{\tilde{Y}}+2T_Y-12g^2_1\bigg)
\end{align}

\bibliographystyle{utphys}
\bibliography{reference}

\end{document}